\documentclass[%
 reprint,
groupedaddress,
showpacs,%preprintnumbers,
 amsmath,amssymb,
 aps,
prl,
]{revtex4-1}

\usepackage{graphicx}% Include figure files
\usepackage{dcolumn}% Align table columns on decimal point
\usepackage{bm}% bold math
\begin{document}

\newcommand{\be}{\begin{equation}}
\newcommand{\ee}[1]{\label{#1}\end{equation}}
\newcommand{\bem}{\begin{eqnarray}}
\newcommand{\eem}[1]{\label{#1}\end{eqnarray}}
\newcommand{\eq}[1]{Eq.~(\ref{#1})}
\newcommand{\Eq}[1]{Equation~(\ref{#1})}
\newcommand{\vp}[2]{[\mathbf{#1} \times \mathbf{#2}]}

%\preprint{APS/123-QED}

\title{Comment on ``Amplitude of waves in the Kelvin-wave cascade''}% Force line breaks with \\

%\tableofcontents

\author{E.  B. Sonin}
 \affiliation{Racah Institute of Physics, Hebrew University of
Jerusalem, Givat Ram, Jerusalem 9190401, Israel}%Lines break automatically or can be forced with \\

\date{\today}% It is always \today, today,

\begin{abstract}
        \end{abstract}

 \maketitle

\citet{EltLvo} derived the relation between the amplitude of Kelvin waves and the energy flux in the Kelvin-wave cascade.   This  returns us to the rather old, but still unresolved dispute on the role  of the tilt symmetry and the locality  in the Kelvin-wave cascade (see Sec.~14.6 of the book \cite{EBS} for references).

\citet{Koz} investigated the Kelvin wave cascade using the Boltzmann equation for the Kelvin modes. They took into account the weak  6-waves interaction and used the locality condition similar to that in the classical Kolmogorov cascade: the energy flux in the space of wave numbers  $k$ depends only on the energy density at  $k$ of the same order of magnitude.  \citet{Lvo} challenged their analysis arguing that  the cascade is connected to the 4-wave interaction despite the latter breaks the rotational invariance and does depend on the tilt of the vortex line with respect to some direction. In the general case of the $n$-wave interaction the expression connecting the energy flux $\epsilon$ in the $k$ space and the the energy density $E_k$ is \cite{SonKW,EBS}
\be
E_k \sim \kappa^2\Lambda \left( \epsilon \over \kappa^3\right)^{1\over n-1} k ^{-{n+1\over n-1}}.
   \ee{spect}
Here $\kappa $ is the circulation quantum  and $\Lambda =\ln {\ell \over a_0}$ is the large logarithm, which depends on the ratio of the intervortex distance or the vortex line curvature radius $\ell$ and the vortex core radius $a_0$. We use notations of \citet{EltLvo}  and their energy normalization.  Here and further on we ignore all numerical factors in our expressions as not important for our qualitative analysis.      

At $n=6$ \eq{spect} gives the spectrum $E_k \propto k^{-7/5}$ of \citet{Koz}, while at $n=4$ one obtains
\be
E_k \sim \kappa^2\Lambda \left( \epsilon \over \kappa^3\right)^{1\over 3} k ^{-5/3}.
   \ee{LN}
This agrees with the spectrum $E_k \propto k^{-5/3}$ of \citet{Lvo}.

However, L'vov and Nazarenko denied not only symmetry arguments, but also the assumption of locality. Since they believed that the Kelvin mode-mode interaction {\em must} depend on the tilt of the vortex line,  they concluded that the interaction vertices in the Boltzmann equation are determined by divergent integrals and the locality assumption is invalid.  Meanwhile, \eq{spect}, as well as its particular case \eq{LN}, was derived assuming locality.  Instead of   \eq{LN},  the nonlocal scenario of L'vov and Nazarenko yields \cite{Lvo11}  
\be
E_k \sim {\kappa^2\Lambda\over \Psi^{2/3}} \left( \epsilon \over \kappa^3\right)^{1\over 3} k ^{-5/3}.
   \ee{LNn}
Here the dimensionless parameter
\be
\Psi\sim {1\over \Lambda \kappa^2} \int\limits_{k_{min}}^\infty E_k\,dk
           \ee{psi}
takes into account the effect of nonlocality since it is an integral over the whole Kelvin-wave cascade interval in  the $k$  space. The lower border of this interval is $k_{min}$.
From Eqs.~(\ref{LNn}) and (\ref{psi}) one obtains that
\be
\Psi\sim \left( \epsilon \over \kappa^3 k_{min}^2 \right)^{1\over 5}.
           \ee{psiE}
So the nonlocality does not affect the dependence on $k$ but does change the dependence on the energy flux $\epsilon$. 

The outcome of the nonlocal scenario is not clear without an evaluation of the minimal wave number $k_{min}$. In the theory of  quantum turbulence $k_{min}$ is the wave number $\sqrt{\cal L}$, at which the crossover from the classical Kolmogorov cascade to the Kelvin-wave cascade occurs.  Here ${\cal L}$ is the vortex line length per unit volume  in Vinen's theory of the 3D vortex tangle.  On the other hand, in agreement with \citet{EltLvo},  the parameter $\Psi$ determines also the ratio of the vortex line length increased by the Kelvin waves participating in the cascade to the length of the straight vortex in the ground state. 
The crossover is determined by the condition that this ratio is on the order of unity \cite{EBS}. If  $\Psi \sim 1$  the energy density \eq{LNn} obtained from the nonlocal scenario does not differ  from the energy density \eq{LN} derived under the assumption of locality. Maybe the reason for the insensitivity of the Kelvin-wave cascade to nonlocal effects deserves a further investigation, but at least it is premature to discard the locality assumption as physically irrelevant.

Also I would like to comment  the statement of \citet{EltLvo} that ``finally the L'vov--Nazarenko model got supported by numerical simulations \cite{Krs,BL}''. This can be interpreted as persisting on the previous claims of the proponents of the L'vov--Nazarenko scenario that the scenario is universal despite it breaks the tilt symmetry. The author of the present comment thinks that it is a bad idea to check the laws of symmetry experimentally or by numerical simulations. If an experiment or a numerical simulation is in conflict with the symmetry law (e.g., the energy conservation law based also on symmetry) the experiment or the simulation must be  reconsidered, but not the other way around.
Suppose that there is a theory based on the crystal cubic symmetry, but they do experiments with parallelepiped samples. Disagreement with the original theory  does not mean that the theory must be discarded. This does mean that one should do experiments at the conditions when the sample shape is not important, e.g. at spatial scales much less than the sample size. In numerical simulations  \cite{Krs,BL} the tilt symmetry was broken since the simulations dealt with the vortex stretched between two parallel surfaces. Probably the spectrum compatible with the tilt invariance could be observed at shorter scales (larger wave numbers $k$). 

In contrast to the case  of the 3D vortex tangle, in numerical simulations of the Kelvin-wave cascade in a straight vortex, the condition $\Psi \sim 1$ is not obligatory, since $k_{min}$ and the energy flux $\epsilon$ can be chosen independently. All scenarios of the Kelvin-wave cascade discussed  above used the theory of weak turbulence valid strictly speaking only if $\Psi \ll 1$. If $\Psi \gg 1$ the turbulence is strong and the spectrum is given by \eq{spect} at $n\to \infty$. This is the spectrum $E_k \sim 1/k$ predicted by \citet{Vin3}.  However, the condition $\Psi \sim 1$  should be imposed on the simulation parameters if one wants to reach  better imitation of processes in the  3D vortex tangle.

In summary: (i)The analysis of \citet{EltLvo} demonstrates that the possible nonlocality of the energy flux in the Kelvin-wave cascade has no essential effect on the Kelvin-wave cascade   in the 3D vortex tangle expected by L'vov and Nazarenko.
(ii)There is no conflict between the Kozik--Svistunov and the L'vov--Nazarenko scenarios. They are valid for different external conditions.

I thank Vladimir Eltsov and Victor L'vov for useful remarks.

%\bibliography{bibBook}



\end{document}